\documentclass[english,prd,twocolumn,nofootinbib,preprintnumbers]{revtex4}

\usepackage[latin1]{inputenc}
\usepackage{graphicx}
\usepackage{amsmath,amsthm,amssymb}
\usepackage{enumitem}

\usepackage{dcolumn}
\usepackage{hyperref}
\usepackage{bm}

\def\be{\begin{equation}}
\def\ee{\end{equation}}
\def\ba{\begin{eqnarray}}
\def\ea{\end{eqnarray}}
\def\bs{\begin{subequations}}
\def\es{\end{subequations}}

\usepackage{color}

\begin{document}

\title{Cosmology of a covariant Galileon field}

\author{Antonio De Felice}
\affiliation{Department of Physics, Faculty of Science, Tokyo University of Science, 
1-3, Kagurazaka, Shinjuku-ku, Tokyo 162-8601, Japan}

\author{Shinji Tsujikawa}
\affiliation{Department of Physics, Faculty of Science, Tokyo University of Science, 
1-3, Kagurazaka, Shinjuku-ku, Tokyo 162-8601, Japan}

\begin{abstract}

We study the cosmology of a covariant scalar field
respecting a Galilean symmetry in flat space-time.
We show the existence of a tracker solution that finally 
approaches a de Sitter fixed point responsible for 
cosmic acceleration today. 
The viable region of model parameters
is clarified by deriving conditions under which
ghosts and Laplacian instabilities of scalar and tensor 
perturbations are absent.
The field equation of state exhibits a peculiar phantom-like 
behavior along the tracker, which allows a possibility 
to observationally distinguish the Galileon gravity from the 
$\Lambda$CDM model.

\end{abstract}

\date{\today}

\maketitle

The problem of dark energy responsible for cosmic acceleration today has 
motivated the idea that the gravitational law may be modified from General 
Relativity (GR) on large scales (see Refs.~\cite{review} for reviews). 
On the other hand, one needs to recover 
Newton gravity at short distances for the compatibility with 
solar-system experiments.   
Besides the chameleon mechanism \cite{chameleon} 
based on the density-dependent matter coupling with a scalar field
(used also in $f(R)$ theories \cite{fRchame,scalaron}), there is
another way to recover GR at short distances: the Vainshtein 
mechanism \cite{Vain} based on non-linear field self-interactions such as
$\square \phi (\nabla \phi)^2$, where 
$(\nabla \phi)^2 \equiv \partial^{\mu} \phi \partial_{\mu}\phi$.
This non-linear effect has been employed for the brane-bending mode 
of the self-accelerating branch in the Dvali-Gabadadze-Porrati (DGP) 
braneworld \cite{DGP}, but the DGP model 
is unfortunately plagued by a ghost problem \cite{DGPghost}.

In order to avoid the appearance of ghosts, it is important to keep the 
field equations up to second-order in time-derivatives.
A scalar field $\phi$ called ``Galileon'' \cite{Nicolis}, whose action
is invariant under the Galilean symmetry 
$\partial_\mu\phi\to\partial_\mu\phi+b_\mu$ 
in flat space-time, allows five field 
Lagrangians that give rise to derivatives up to second-order 
(see Refs.~\cite{DeffaGal}-\cite{Sami} for related works).
If the analysis in \cite{Nicolis} is extended to the curved 
space-time, one needs to introduce couplings between the field 
$\phi$ and the curvature tensors for constructing the 
Lagrangians free from higher-order derivatives
in the equations of motion \cite{DeffaGal}.

The five covariant Lagrangians that respect the Galilean 
symmetry in flat space-time are given by 
\begin{eqnarray}
& & {\cal L}_1=M^3 \phi\,,\quad 
{\cal L}_2=(\nabla \phi)^2\,,\quad
{\cal L}_3=(\square \phi) (\nabla \phi)^2/M^3\,, \nonumber \\
& & {\cal L}_4=(\nabla \phi)^2 \left[2 (\square \phi)^2
-2 \phi_{;\mu \nu} \phi^{;\mu \nu}-R(\nabla \phi)^2/2 \right]/M^6,
\nonumber \\
& & {\cal L}_5=(\nabla \phi)^2 [ (\square \phi)^3
-3(\square \phi)\,\phi_{; \mu \nu} \phi^{;\mu \nu} \nonumber \\
& &~~~~~~~+2{\phi_{;\mu}}^{\nu} {\phi_{;\nu}}^{\rho}
{\phi_{;\rho}}^{\mu} 
-6 \phi_{;\mu} \phi^{;\mu \nu}\phi^{;\rho}G_{\nu \rho} ]
/M^9\,,
\label{lag}
\end{eqnarray}
where a semicolon represents a covariant derivative, $M$ is a constant 
having a dimension of mass, and $G_{\nu \rho}$ is the Einstein tensor.
In this Letter we study the cosmology based on the action 
\begin{equation}
S=\int {\rm d}^4 x \sqrt{-g}\,\left[ \frac{M_{\rm pl}^2}{2}R+
\frac12 \sum_{i=1}^5 c_i {\cal L}_i \right]
+\int {\rm d}^4 x\, {\cal L}_{M}\,,
\label{action}
\end{equation}
where $M_{\rm pl}$ is the reduced Planck mass and 
$c_i$ are constants. 
For the matter Lagrangian ${\cal L}_{M}$
we take into account perfect fluids of 
radiation (density $\rho_r$) and 
non-relativistic matter (density $\rho_m$).
Although the cosmological dynamics up to ${\cal L}_4$
were discussed in Ref.~\cite{Sami}, we will show that 
inclusion of ${\cal L}_5$ is crucially important to determine
the full Galileon dynamics. Moreover the viable 
parameter space will be clarified for such full theory.

In the flat Friedmann-Lema\^{i}tre-Robertson-Walker (FLRW) Universe
with a scale factor $a(t)$, the variation of the action (\ref{action}) 
leads to the following equations of motion
\begin{eqnarray}
& & 3M_{\rm pl}^2 H^2=\rho_{\rm DE}+\rho_m+\rho_r\,,
\label{basic1} \\
& & 3M_{\rm pl}^2 H^2+2M_{\rm pl}^2 \dot{H}=-P_{\rm DE}
-\rho_r/3\,,
\label{basic2}
\end{eqnarray}
where $H=\dot{a}/a$ is the Hubble parameter (a dot represents 
a derivative with respect to cosmic time $t$), and 
\begin{eqnarray}
\rho_{\rm DE} &\equiv& -c_1 M^3 \phi/2-c_2 \dot{\phi}^2/2
+3c_3 H \dot{\phi}^3/M^3 \nonumber \\
& &-45 c_4 H^2 \dot{\phi}^4/(2M^6)
+21c_5 H^3 \dot{\phi}^5/M^9,\\
P_{\rm DE} &\equiv&  c_1 M^3 \phi/2-c_2 \dot{\phi}^2/2
-c_3 \dot{\phi}^2 \ddot{\phi}/M^3 \nonumber \\
& &+3c_4 \dot{\phi}^3 [8H\ddot{\phi} +(3H^2+2\dot{H})
\dot{\phi}]/(2 M^6) \nonumber \\
& & -3c_5 H \dot{\phi}^4 [5H \ddot{\phi}+2(H^2+\dot{H}) 
\dot{\phi} ]/M^9\,.
\end{eqnarray}
The matter fluids obey the continuity equations 
$\dot{\rho}_m+3H \rho_m=0$ and 
$\dot{\rho}_r+4H \rho_r=0$.
{}From Eqs.~(\ref{basic1}) and (\ref{basic2})
the dark component also satisfies 
$\dot{\rho}_{\rm DE}+3H(\rho_{\rm DE}+P_{\rm DE})=0$.
We define the dark energy equation of state $w_{\rm DE}$
and the effective equation of state $w_{\rm eff}$, as 
$w_{\rm DE} \equiv P_{\rm DE}/\rho_{\rm DE}$ and
$w_{\rm eff} \equiv -1-2\dot{H}/(3H^2)$.
The latter is known by the background expansion history 
of the Universe.

Since we are interested in the case where the late-time cosmic 
acceleration is driven by field kinetic terms without a potential, 
we set $c_1=0$ in the following discussion.
When $c_1=0$ and $c_2 \neq 0$ the only solution 
to Eqs.~(\ref{basic1}) and (\ref{basic2}) in the 
Minkowski background ($H=0$)
without matter corresponds to $\dot{\phi}=0$.
We introduce the following quantities useful to describe 
the cosmological dynamics
\begin{equation}
r_1 \equiv \dot{\phi}_{\rm dS}H_{\rm dS}/(\dot{\phi}H)\,,
\qquad
r_2 \equiv (\dot{\phi}/\dot{\phi}_{\rm dS})^4/r_1\,,
\label{rdef}
\end{equation}
where $\dot{\phi}_{\rm dS}$ and  $H_{\rm dS} \approx 10^{-60}\,M_{\rm pl}$
are the field velocity and the Hubble parameter at the 
de Sitter (dS) solution, respectively.
The mass $M$ is related to $H_{\rm dS}$ via 
$M^3=M_{\rm pl}H_{\rm dS}^2$.
At the dS point one has $r_1=1$ 
and $r_2=1$.
Equation (\ref{basic1}) can be written as 
$\Omega_{m}+\Omega_r+\Omega_{\rm DE}=1$, where
$\Omega_m=\rho_m/(3M_{\rm pl}^2H^2)$, 
$\Omega_r=\rho_r/(3M_{\rm pl}^2H^2)$, and 
\begin{equation}
\Omega_{\rm{DE}} = -\frac16 c_2 x_{\rm dS}^2 r_1^3 r_2
+c_3 x_{\rm dS}^3 r_1^2 r_2 -\frac{15}{2} c_4 x_{\rm dS}^4 r_1 r_2
+7c_5x_{\rm dS}^5 r_2,
\label{OmeDE}
\end{equation}
where $x_{\rm dS} \equiv \dot{\phi}_{\rm dS}/(H_{\rm dS}M_{\rm pl})$.
Since $\Omega_{\rm DE}=1$ at the dS point, 
Eq.~(\ref{OmeDE}) gives a relation between the terms
$c_2 x_{\rm dS}^2$, $c_3 x_{\rm dS}^3$, 
$\alpha \equiv c_4 x_{\rm dS}^4$,
$\beta \equiv c_5x_{\rm dS}^5$.
Combining this with another relation coming from 
Eq.~(\ref{basic2}), we obtain
\begin{equation}
c_2 x_{\rm dS}^2=6+9\alpha-12\beta\,,\quad
c_3 x_{\rm dS}^3=2+9\alpha-9\beta\,.
\label{cre}
\end{equation}
It is useful to use $\alpha$ and $\beta$ because
the coefficients of physical quantities (such as $\Omega_{\rm DE}$)
can be expressed in terms of those quantities 
thanks to Eq.~(\ref{cre}). The relations (\ref{cre}) are not subject to change under a rescaling $c_i\to c_i/\gamma^i$ and $x_{\rm dS}\to\gamma x_{\rm dS}$, where $\gamma$ is a real number. Therefore rescaled choices of $c_i$ will lead to the same dynamics (as they have the same $\alpha$ and $\beta$) for both the background and 
the linear perturbation, which implies that redefining the coefficients 
$c_i$ in terms of $\alpha$ and $\beta$ is convenient.

The autonomous equations for the variables $r_1, r_2, \Omega_r$
follow from Eqs.~(\ref{basic1}), (\ref{basic2}), and fluid equations.
One can show that there is an equilibrium point characterized by 
\begin{equation}
r_1=1\,, \quad {\rm i.e.} \quad 
\dot{\phi}H={\rm constant}\,,
\label{r11}
\end{equation}
at which the variables $r_2$ and $\Omega_r$ satisfy
\begin{equation}
\label{eq:dR2}
\hspace{-0.25cm}
r_2' =\frac{2 r_2 \left( 3-3r_2+\Omega _r\right)}
{1+r_2},\quad
\Omega_r' =\frac{\Omega _r
\left(\Omega _r-1-7 r_2\right)}{1+r_2},
\end{equation}
where a prime represents a derivative with respect to $N=\ln a$.
This result is interesting because it shows the universality of the
equations of motion without 
any dependence on $\alpha$ and $\beta$.
Along the solution (\ref{r11}), the field velocity evolves as 
$\dot{\phi} \propto t$ during 
radiation and matter eras ($H \propto 1/t$). 
There is also a simple relation
$\Omega_{\rm DE}=r_2$ along the solution $r_1=1$.

We have three fixed points: 
(a) $(r_1, r_2, \Omega_r)=(1,0,1)$ [radiation],
(b) $(r_1, r_2, \Omega_r)=(1,0,0)$ [matter],
(c) $(r_1, r_2, \Omega_r)=(1,1,0)$ [dS].
The stability of these points can be analyzed by considering 
linear perturbations $\delta r_1, \delta r_2, \delta \Omega_r$
about them. The perturbation $\delta r_1$ satisfies
\begin{equation}
\delta r_1'=-\frac{9+\Omega_r+3r_2}
{2(1+r_2)} \delta r_1\,,
\end{equation}
which shows that, in the regime $0 \le r_2 \le 1$ and 
$\Omega_r \ge 0$, the solution is stable in the direction of $r_1$. 
Since the dS point is stable in the 
other two directions, the solutions
finally approach it. The points (a) and (b) are saddle because
they are unstable in the direction of $r_2$.

Along the solution (\ref{r11}) we have 
$\rho_{\rm DE}=3M^6/H^2$, 
$P_{\rm DE}=-3M^6 (2+w_{\rm eff})/H^2$, and 
\begin{equation}
\hspace{-0.2cm}
w_{\rm DE}=-2-w_{\rm eff}
=-\frac{\Omega _r+6}{3\,(r_2+1)},\quad
w_{\rm eff}=\frac{\Omega _r-6\, r_2}
{3 \left(r_2+1\right)}.
\label{wana1}
\end{equation}
{}From the radiation era to the dS epoch 
the effective equation of state evolves as
$w_{\rm eff}=1/3 \to 0  \to -1$, 
whereas the dark energy equation of state exhibits
a peculiar evolution: $w_{\rm DE}=-7/3 \to -2 \to -1$. 

The evolution of dark energy is different depending on 
the initial conditions of $(r_1, r_2, \Omega_r)$.
If they are chosen to be close to the fixed point (a) 
at the onset of the radiation era, then
the solutions follow the sequence (a) $\to$ (b) $\to$ (c).
If $r_1 \ll 1$ initially, the dominant contribution to $\Omega_{\rm DE}$
comes from the term ${\cal L}_5$, i.e. 
$\Omega_{\rm DE} \simeq 7\beta r_2$.
In this case the solutions approach $r_1=1$
at late times with the increase of $r_1$.
For the initial conditions with $r_1 \gg 1$ the term ${\cal L}_2$ gives 
the dominant contribution to $\Omega_{\rm DE}$, but this case
is not viable because the field kinetic energy decreases rapidly 
as in quintessence without a potential.
Numerical simulations show that if $r_1 \lesssim 2$ initially 
the solutions approach $r_1=1$, but in the opposite case 
the Universe finally reaches the matter-dominated epoch.

Let us find the allowed parameter space in terms of $(\alpha, \beta)$
by deriving the conditions for the avoidance of ghosts and 
instabilities of scalar and tensor perturbations. 
Using the Faddeev-Jackiw method \cite{Faddeev}, 
the action (\ref{action}) can be expanded at second-order
in perturbations. Following the similar procedure to 
that given in Refs.~\cite{Suyama}, 
the no-ghost condition for the scalar sector of the action 
(\ref{action}) is given by 
\begin{equation}
Q_S \equiv -s/(1+\mu_3)^2>0\,,
\label{Qsdef}
\end{equation}
where $s \equiv 6(1+\mu_1)(\mu_1+\mu_2+\mu_1 \mu_2
-2\mu_3-\mu_3^2)$, and 
\begin{eqnarray}
\hspace{-1.1cm} 
& & \mu_1 \equiv 3\alpha r_1 r_2/2-3\beta r_2\,,\\
\hspace{-1.1cm} 
& & \mu_2 \equiv (3\alpha-4\beta+2)r_1^3 r_2/2
-2(9\alpha-9\beta+2)r_1^2 r_2 \nonumber \\
\hspace{-1.1cm} 
& &~~~~~~~+45\alpha r_1 r_2/2-28\beta r_2\,,\\
\hspace{-1.1cm} 
& & \mu_3 \equiv -(9\alpha-9\beta+2)r_1^2 r_2/2+
15\alpha r_1 r_2/2 -21 \beta r_2/2.
\end{eqnarray}
The condition for the avoidance of Laplacian instabilities
associated with the scalar field propagation speed is  
\begin{eqnarray}
& &\hspace{-0.4cm}
c_S^2=\{(1+\mu_1)^2 [2\mu_3'-(1+\mu_3)(5+3w_{\rm eff})
+3\Omega_m+4\Omega_r] \nonumber \\
& & \hspace{-0.4cm} -4\mu_1' (1+\mu_1)(1+\mu_2)
+2(1+\mu_3)^2 (1+\mu_4)\}/s>0,
\label{csdef}
\end{eqnarray}
where 
\begin{equation}
\mu_4 \equiv -\alpha r_1 r_2/2+3\beta r_2
(3+3w_{\rm eff}-3r_1'/r_1-r_2'/r_2)/2\,.
\end{equation}
Similar calculations for the tensor perturbation 
lead to 
\begin{eqnarray}
\label{eq:QT}
& & Q_T \equiv 3\alpha r_1 r_2 /4-3\beta r_2/2
+1/2 > 0 \,, \\
\label{eq:cT}
& & c_T^2=\frac{2 r_1\left( 2- \alpha r_1 r_2 \right)
-3 \beta  \left(r_2 r_1'+r_1r_2'\right)}{2 r_1\left(2+3 \alpha r_1 r_2
-6 \beta  r_2 \right)}>0\,.
\end{eqnarray}
\begin{figure}
\includegraphics[height=2.4in,width=3.2in]{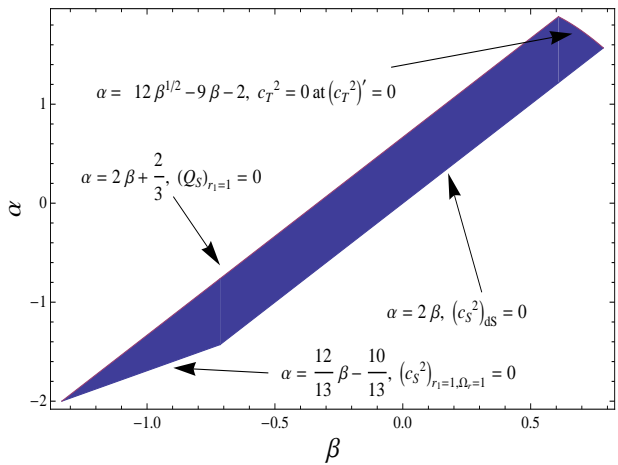}
\caption{\label{fig1} The viable parameter space in the 
$(\alpha, \beta)$ plane for the branch $r_2>0$.
We also show several conditions that determine the border
between the allowed and excluded regions.}
\end{figure}

We consider the following three different regimes.
\begin{itemize}
\item (i) $r_1 \ll 1$, $r_2 \ll 1$

This characterizes the early cosmological epoch in which 
the term ${\cal L}_5$ dominates the dynamics of the field. 
For the scalar modes we have 
$Q_S \simeq 60 \beta r_2$ and $c_S^2 \simeq (1+\Omega_r)/40$.
The sign change of $r_2$ implies the appearance of ghosts.
For the initial conditions with $r_2>0$, it is
required that $\beta>0$. 
The Laplacian instabilities of the scalar modes can be avoided
because $c_S^2 \simeq 1/20$ and $c_S^2 \simeq 1/40$ 
during radiation and matter eras, respectively.
Since $Q_T \simeq 1/2$ and $c_T^2\simeq1+3\beta
r_2(5-3\Omega_r)/8 \simeq 1$, 
the tensor modes do not provide additional constraints.
We also have 
\begin{equation}
w_{\rm DE} \simeq -(1+\Omega_r)/8\,,\qquad
w_{\rm eff} \simeq \Omega_r/3\,,
\label{wana2}
\end{equation}
which is valid for $\Omega_r \gg \{r_1, r_2\}$.
\item (ii) $r_1=1$, $r_2 \ll 1$

This corresponds to the equilibrium point (\ref{r11}) during 
radiation or matter domination. The conditions (\ref{Qsdef}) 
and (\ref{csdef}) reduce to 
\begin{eqnarray}
& & Q_S \simeq 3(2-3\alpha+6\beta)r_2>0\,,
\label{Qscon2} \\
& & c_S^2 \simeq \frac{8+10\alpha-9\beta+
\Omega_r (2+3\alpha-3\beta)}
{3(2-3\alpha+6\beta)}>0\,.
\label{cscon2}
\end{eqnarray}
For the branch $r_2>0$ the first condition 
reduces to $2-3\alpha+6\beta>0$. 
For the tensor modes, we have 
$c_T^2\simeq 1-r_2(4\alpha+3\beta+3\beta\Omega_r)/2 \simeq 1$
and $Q_T>0$.
\item (iii) $r_1=1$, $r_2=1$

This corresponds to the dS point, at which the 
conditions (\ref{Qsdef}), (\ref{eq:QT}), (\ref{csdef}), 
and (\ref{eq:cT}) are given by 
\begin{eqnarray}
& & Q_S = \frac{4-9(\alpha-2\beta)^2}{3(\alpha-2\beta)^2}>0 \,,
\label{Qscon3} \\
&& Q_T=(2+3 \alpha -6 \beta )/4>0\,,\label{QTcon3}\\
& & c_S^2 = \frac{(\alpha-2\beta)
(4+15\alpha^2-48\alpha \beta +36\beta^2)}
{2[4-9(\alpha-2\beta)^2]}>0,
\label{cscon3}\\
&& c_T^2=\frac{2-\alpha }{2+3 \alpha -6 \beta }>0\, .
\label{cTcon3}
\end{eqnarray}
\end{itemize}
If $\beta>0$, $c_T^2$ can have a minimum
during the transition from the regime $r_2 \ll 1$ to $r_2 \simeq 1$. 
This value tends to decrease as $\beta$ approaches 1.
Imposing that $c_T^2>0$ at the minimum, we obtain  
$\alpha<12\sqrt\beta-9\beta-2$. 
In Fig.~\ref{fig1} we illustrate the region in which this condition 
as well as the conditions (\ref{Qscon2})-(\ref{cTcon3}) are 
satisfied for $r_2>0$. 
Numerical simulations confirm that for the parameters inside 
the shaded region in Fig.~\ref{fig1} the no-ghost and 
stability conditions are not violated even in 
the intermediate cosmological epoch.

\begin{figure}[t]
\includegraphics[height=2.5in,width=3.4in]{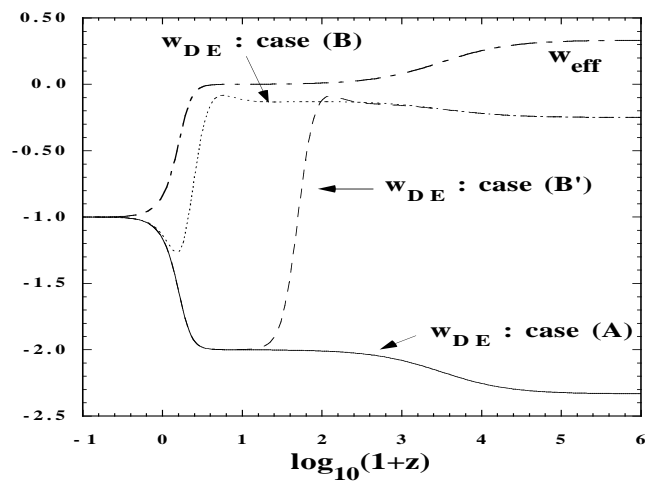}
\caption{\label{fig2} Evolution of $w_{\rm eff}$ and $w_{\rm DE}$ 
for the cases: (A) $\alpha=-1.4$, $\beta=-0.8$, $x_{\rm dS}=1$ 
with initial conditions $r_1=1$, $r_2=10^{-60}$, 
$\Omega_r=0.99999$ at the redshift $z=3.11 \times 10^8$, 
(B) $\alpha=0.1$, $\beta=0.049$, 
$x_{\rm dS}=1$ with initial conditions $r_1=5 \times 10^{-11}$, 
$r_2=8 \times 10^{-12}$, $\Omega_r=0.999995$ at 
$z=6.44 \times 10^8$, and
(B$'$) the same $\alpha$, $\beta$, $x_{\rm dS}$ as in the case (B)
but with different initial conditions $r_1=5 \times 10^{-7}$, 
$r_2=8 \times 10^{-16}$, $\Omega_r=0.9995$ at 
$z=6.72 \times 10^6$.}
\end{figure}
In Fig.~\ref{fig2} we plot the variation of $w_{\rm DE}$ and 
$w_{\rm eff}$ versus the redshift $z$
for several different model parameters and initial conditions. 
In the case (A) the initial condition is chosen to be $r_1=1$, so that 
$w_{\rm DE}$ and $w_{\rm eff}$ evolve according to Eq.~(\ref{wana1}) 
with the variation of $r_2$ and $\Omega_r$.
While the evolution of $w_{\rm eff}$ is similar to that 
in the $\Lambda$CDM model, the dark energy equation of state 
evolves from the regime $w_{\rm DE}<-1$ to the dS attractor with 
$w_{\rm DE}=-1$. 
The cases (B) and (B$'$) in Fig.~\ref{fig2} correspond to 
the initial conditions in the regime (i). As estimated by Eq.~(\ref{wana2}), 
$w_{\rm DE}$ starts to evolve from $-1/4$ and 
reaches the value $-1/8$ during the matter era.
The evolution of $w_{\rm DE}$ is different depending on 
the epoch at which $r_1$ grows to the order of 1. 
In the case (B) the solutions reach the regime 
$r_1 \sim 1$ only recently, whereas in the case (B$'$)
the approach of this regime occurs much earlier.
The equilibrium point (\ref{r11}) can be regarded as
a tracker that attracts solutions with different initial 
conditions to a common trajectory.
Before approaching the tracker, the solutions cross 
the boundary $w_{\rm DE}=-1$ without any unstable
behavior of perturbations.
Note that there are no significant differences for the
variation of $w_{\rm eff}$ between the cases 
(A) and (B) [also (B$'$)].

\begin{figure}
\includegraphics[height=2.5in,width=3.5in]{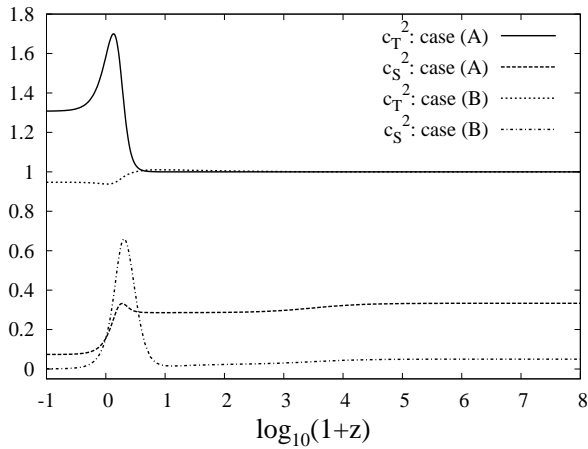}
\caption{\label{fig3} Evolution of $c_S^2$ and $c_T^2$ for the cases (A) and 
(B) as in Fig.~\ref{fig2}.}
\end{figure}

In Fig.~\ref{fig3} we show the evolution of $c_S^2$ and $c_T^2$ for
the same model parameters and initial conditions as in
Fig.~\ref{fig2}.  In the case (A) the scalar propagation speed remains
sub-luminal, as estimated by Eqs.~(\ref{cscon2}) and (\ref{cscon3}).
For $\alpha=-1.4$ and $\beta=-0.8$, Eq.~(\ref{cTcon3}) shows that at
the dS point the tensor mode becomes super-luminal.  However,
both the scalar and tensor modes can be sub-luminal at the dS point, 
as in the case (B)
of Fig.~\ref{fig3} ($\alpha=0.1$, $\beta=0.049$).

For the initial conditions starting from the regime (i) we 
require $\beta>0$ to avoid ghosts.
Under the conditions (\ref{Qscon2}), (\ref{Qscon3}), 
and (\ref{cscon3}) with $\beta>0$, one can show that $c_S^2$ 
in Eq.~(\ref{cscon2}) gets larger than 1.
If the solutions approach the tracker in the regime (ii) 
long before the dS epoch, there is a period in which $c_S^2$ 
exceeds 1. This super-luminal propagation can be avoided
if $r_1$ grows to the order of unity only recently.
The case (B) in Fig.~\ref{fig3} corresponds to such an example 
for which $c_S^2$ has a peak smaller than 1
after the matter era.
In this case the tensor mode is slightly super-luminal 
in the regime (i). In general there is a period in which 
the propagation speed of either scalar or tensor 
modes exceeds 1.
However, this does not necessarily imply the inconsistency 
of theory because of the possibility for the absence  
of closed causal curves \cite{Sami}.

In summary we have studied the cosmology for the full Galileon action
(\ref{action}) and derived all conditions for the consistency
of such theory. We have shown that, under the conditions (\ref{cre}), 
there exist stable dS solutions responsible for dark energy.
In spite of the complexities of Galileon Lagrangians,
the conditions for the 
avoidance of ghosts and Laplacian instabilities constrain the allowed
parameter space in terms of the variables $\alpha$ and 
$\beta$ in a simple way. 
While the evolution of $w_{\rm DE}$, $c_S^2$ and $c_T^2$ is different 
depending on the model parameters and the initial conditions of $r_1$, 
we have derived convenient analytic formulas to evaluate those quantities
 in three distinct regimes.

There are several interesting applications and generalizations 
of Galileon gravity. First, the study of cosmological 
perturbations may provide some signatures 
for the modification of gravity from GR.
The last term of ${\cal L}_4$ in Eq.~(\ref{lag}), for example, gives rise to 
a correction of the order $\alpha r_1 r_2$ to the bare gravitational 
constant. This affects the effective gravitational coupling $G_{\rm eff}$
that appears in the equation of matter perturbations.
Also it will be possible to constrain the Galileon models from 
the time variation of $G_{\rm eff}$.
Second, the study of spherically symmetric solutions
in both weak and strong gravitational backgrounds can 
allow us to understand how the Vainshtein mechanism works
in general. Third, it is possible to extend the Galileon Lagrangian (\ref{lag})
to the theory in which the field $\phi$ is replaced by a general  
function $f(\phi)$. We expect that such analyses will 
provide us deep insight on the possible modification of gravity and 
that it will shed new light on the nature of dark energy.

\vspace{0.1cm}
{\it Acknowledgements}--We thank JSPS for financial support
(Nos.~09314, 30318802, and 21111006).


\end{document}